# Quantum Annealing-Enhanced Virtual Traffic Lights and its Evaluation Using a Quantum-in-the-Loop Simulation Testbed

Abyad Enan*

Clemson University, Clemson, South Carolina, U.S., aenan@clemson.edu

M Sabbir Salek

Clemson University, Clemson, South Carolina, U.S., msalek@clemson.edu

Mashrur Chowdhury

Clemson University, Clemson, South Carolina, U.S., mac@clemson.edu

Gurcan Comert

North Carolina A&T State University, Greensboro, North Carolina, U.S., gcomert@ncat.edu

Sakib Mahmud Khan

MITRE Corporation, McLean, Virginia, U.S., sakibkhan@mitre.org

Reek Majumder

Clemson University, Clemson, South Carolina, U.S., rmajumd@clemson.edu

Virtual Traffic Light (VTL) is a traffic control method that does not require traffic signal-related infrastructure for roadway intersections. Connected vehicles (CVs) are given right-of-way based on prevailing traffic conditions, such as estimated times of arrival (ETAs) of vehicles, the number of CVs in different approaches, and their emissions. These factors are considered in line with the objectives of the VTL application. Aiming to optimize traffic flow by reducing delays, VTL generates Signal Phase and Timing (SPaT) data for CVs approaching an intersection. Our VTL method considers the delay each CV would cause for other CVs if given the right-of-way. However, the stochastic nature of vehicle arrivals at intersections increases the complexity of the optimization problem, making it challenging for classical computers to determine optimal solutions in real-time. To address this limitation, we develop a VTL method designed to minimize stopped delays for CVs at an intersection by leveraging the efficacies of existing quantum computers that determine the best outcome from all possible combinations. This method employs Quadratic Unconstrained Binary Optimization (QUBO), a mathematical framework commonly used in quantum computing, to formulate the VTL problem as a stopped-delay-minimization challenge. To evaluate our method for roadway traffic with varying traffic volumes, we integrate an open-source microscopic roadway traffic simulator, Simulation for Urban Mobility (SUMO), with a cloud-based D-Wave quantum computer. Our analysis reveals that our quantum computing-supported VTL outperforms the classical optimization-based VTL by significantly reducing stopped delays at intersections and travel time through the roadway sections crossing the intersections.

* Corresponding Author.

## 1 INTRODUCTION

Traffic congestion impacts our daily lives, increasing travel times and emissions [1]. Factors contributing to traffic congestion include construction and accidents [40], and outdated traffic light timing and patterns [18]. According to the Federal Highway Administration, 75% of traffic signals in the U.S. could be improved by updating signal plans and timing to reduce congestion [23]. Researchers worldwide are addressing this daunting issue by developing solutions for various use cases, such as Vehicle Traffic Routing Systems (VTRSs) [49], intelligent and adaptive traffic signal control systems [52], and smart toll collection systems [2]. These solutions assume sensors, computational intelligence, and communication systems to supplement traditional traffic infrastructure [2, 11, 49].

Vehicle-to-Everything (V2X) communication technology enables vehicles to communicate with roadside and cloud infrastructure, pedestrians, bicycles, other vehicles, etc. V2X is vital in connected vehicle (CV) applications [29]. In [13], the authors have demonstrated how CVs can effectively communicate with cloud computers for real-time speed advisories at intersections. CVs broadcast Basic-Safety-Messages (BSMs), which consist of vehicles' states and safety-related information, such as position, speed, acceleration, heading, yaw rate, and brake status [17, 55], which are used for the real-time speed advisory application.

In a CV environment, Virtual Traffic Light (VTL), an emerging technology, holds the potential for addressing traffic congestion at an intersection. VTL is a CV-based traffic control concept that does not require any physical traffic light infrastructure. All CVs at the VTL zones (i.e., the range for each lane of the roadway at an intersection approach in which CVs can be accommodated in a VTL application for optimal traffic flow) are interconnected, and a central server determines which vehicle(s) should pass through the intersection at a given time [54]. The central server could be roadside, backend, or even cloud-based. Vehicles approaching an intersection could be given the right-of-way by the VTL system based on the estimated times of arrival (ETAs) of each vehicle (i.e., on a first-come, first-served basis) and other factors, such as the number of vehicles in a lane (when vehicles from conflicting phases arrive at the intersection at the same time), the approach having the highest number of vehicles gets the right-of-way. The primary goal of VTL is to ensure continuous traffic flow at an intersection, which could help reduce the stopped delay or waiting time for vehicles at the intersection. The vehicles receive Signal Phase and Timing (SPaT) data from a central server, eliminating the need for a traffic signal controller, traffic lights, sensors, and associated wired and wireless communication infrastructure.

To reduce the delays of CVs at intersections, we develop a Quadratic Unconstrained Binary Optimization (QUBO) objective function to formulate the VTL problem as a stopped delay-minimization challenge. QUBO is a mathematical framework used to solve combinatorial optimization problems. In QUBO, any optimization problem is represented as a quadratic problem over binary variables, with the goal of finding the optimal solution for these variables that minimizes or maximizes an objective function. QUBO is commonly used in quantum computing frameworks, particularly in adiabatic quantum computing and quantum annealing, which aim to solve optimization problems by determining a quantum system's ground state.

To evaluate the VTL method developed in this study, we use the Simulation for Urban Mobility (SUMO) traffic simulation package to simulate CVs under varying traffic volumes. The CVs in SUMO are connected in real-time to a cloud-based quantum annealer, D-Wave machine, which minimizes the QUBO objective function to reduce the delays of CVs based on the principles of quantum annealing. According to [41], a quantum annealer can successfully optimize combinatorial optimization problems like QUBO. After the optimization process, the decisions are sent to the CVs in SUMO, allowing the vehicles to navigate the intersection based on their respective right of way, derived from the optimization's output. For comparison, we implement classical optimizers for the same VTL problem to minimize the stopped delays of CVs. In addition, we compare the performance of our method with traditional infrastructure-based actuated traffic signals.

Traditional infrastructure-based traffic control systems also have security concerns, as they are increasingly vulnerable to cyberattacks due to their poor security practices. Many of such systems lack basic cybersecurity features, such as encryption, secure authentication, or access control [9]. They often rely on wireless communication to enable communication between traffic signal controllers, sensors, and central management systems using unencrypted 900 MHz or 5.8 GHz radios and default credentials that are rarely changed, exposing them to unauthorized access [28]. Once compromised, an attacker can manipulate the signal timing or override traffic logic, leading to significant congestion or safety risks without triggering physical failsafe systems such as the Malfunction Management Unit (MMU) [9]. Recently, a significant cybersecurity vulnerability has been identified in a traffic light controller, Intelight X-1, used widely throughout the United States, where a cybersecurity researcher discovered that the traffic light controller lacks authentication, allowing remote attackers to manipulate signal timings and potentially cause widespread traffic chaos, highlighting critical vulnerabilities in outdated infrastructure [30]. In contrast, when the traffic flow of intersections is controlled by a cloud-based platform, especially in the context of CV applications, it is facilitated with tools to ensure confidentiality, integrity, availability, authentication, accountability, and privacy [44]. When this cloud platform is powered by quantum cloud infrastructures, cybersecurity benefits emerge, particularly in terms of secure communication and trusted access control. The quantum computer, supported by the cloud environment, uses Transport Layer Security (TLS) encryption for all communication between clients and quantum processors, ensuring that the data sent for optimization cannot be intercepted or altered in transit. Access to the cloud system is strictly gated through API token-based authentication, enabling fine-grained control over who can submit problems and view results [16]. This architecture ensures that even if the wider network is exposed to threats, the optimization logic and its associated data remain protected within a hardened, encrypted quantum computing interface.

To the best of the authors' knowledge, this study is the first to develop a quantum optimization method for VTL and integrate SUMO with the D-Wave quantum annealer to develop a simulation testbed for a VTL application. The key contributions of this study are as follows:

- Develop a VTL optimization problem to reduce the stopped delay of vehicles at an intersection.
- Develop a generalized VTL optimization problem that considers all possible vehicle movements at an intersection in accordance with the National Electrical Manufacturers Association (NEMA) phasing.
- Formulate the VTL optimization problem mathematically as a QUBO model, designed to run on a quantum annealer for optimization.

- Develop an interactive simulation testbed by integrating an open-source microscopic traffic simulator, SUMO, with the cloud-based D-Wave quantum annealer.

The rest of the paper is structured as follows: Section 2 reviews related studies. Section 3 discusses quantum annealing and the motivation behind adopting quantum computing for VTL. Section 4 explains our VTL methods. Section 5 covers the evaluation strategies. Section 6 presents the evaluation outcomes. Finally, conclusions are drawn in Section 7.

## 2 RELATED STUDIES

In this section, we review existing work that focuses on infrastructure-less traffic control, particularly virtual traffic lights (VTL). Based on this review, we discuss the limitations and research gaps that our study aims to address.

The authors in [21] presented an infrastructure-less, self-organized traffic control scheme for isolated intersections, which improves traffic mobility by reducing average commute times by over 30% compared to signalized intersections. In their study, they outlined guidelines and rules for how vehicles can communicate and collectively decide traffic flow conditions at intersections. However, the method is not generalized and does not address how it would function for intersections with multi-lane roadways. Similarly, the authors of [6] developed a distributed VTL algorithm leveraging short-range vehicle-to-vehicle (V2V) communication. The authors tested their VTL method in a real-world setting using communication radios that follow the IEEE 802.11p protocol. In their work, the authors provided generalized rules for traffic movements at an intersection, which accounted for various traffic conditions, such as the presence, position, and number of vehicles at different intersection approaches. However, their method is only applicable to intersections with two-lane roadways (one lane in each direction) and does not extend to those with multi-lane configurations. In summary, these studies in [6, 21] primarily focus on VTL systems that rely on predefined rules and conditions. However, in real-world scenarios, vehicle arrivals at intersections are stochastic.

In [29], the authors developed a cooperative vehicle intersection control algorithm for CVs, aiming to reduce vehicle stopped delays, travel times, and fuel consumption. While their method is applicable to intersections with dedicated left-turn and through-movement lanes, they validated its performance only for scenarios involving four through movements. Additionally, their method relies on three different optimization processes running in parallel, which could introduce significant computational load.

A priority-based infrastructure-less vehicle scheduling strategy for autonomous vehicles is presented by [54], where the authors considered Dedicated Short-Range Communication (DSRC) technology for the viability of an infrastructure-less traffic control model for two vehicles through field tests. However, if an intersection timing design problem turns into an NP-hard problem due to considering concurrent movements for vehicles from multiple lanes, similar to other scheduling problems where the computation time increases exponentially with the problem size, it becomes challenging for a classical computer to solve it in a finite time [34]. Quantum computers, such as quantum annealers, can overcome the limitation of solving an NP-hard problem by utilizing the tunneling effect, as explained in the last section [14].

Apart from the development of infrastructure-less traffic control systems such as Virtual Traffic Lights (VTL), recent advancements in quantum annealers have enabled several studies to explore the optimization of traffic signals using this technology. In [26] and [25], the authors proposed methods for optimizing traffic signals through quantum annealing. The authors of [26] presented a global traffic signal control approach where the road network of a city is modeled as a grid, with traffic signals positioned at the intersections. Their method outperforms traditional local controller-based systems. However, the vehicle movements in their model are limited to either horizontal or vertical directions (north-south through or east-west through), which does not fully capture real-world traffic scenarios, such as left turns, right turns, and through movements in both directions of the grid. In [25], the authors also proposed a global intersection control strategy using quantum annealing and considered six distinct vehicle movement patterns at intersections. They demonstrated that their method outperforms traditional fixed-time signal control. Nonetheless, they omitted two additional critical movement types: north–south left turns and east–west left turns (commonly corresponding to right turns in European traffic systems). Moreover, their evaluation, which used a custom simulator with 4,320 randomly positioned vehicles, does not provide insights into how the method would perform under varying traffic volumes or incorporate realistic driving behaviors and models typically captured by standard traffic simulators.

Our study addresses the limitations of existing VTL research by focusing on intersections with multiple lanes in each approach, including dedicated lanes for left-turning vehicles. Our VTL method is presented as a stopped delay minimization problem for CVs to optimize traffic control at intersections, making it adaptable to any scenario. In addition, we evaluate our method under varying traffic volumes and the length of VTL zones, demonstrating its robustness across diverse traffic conditions and intersection configurations

## 3 PRELIMINARIES

Vehicles arriving at an intersection are a dynamic process. Optimizing their movements to reduce their waiting time at an intersection is a non-convex problem. To determine the most optimized traffic control for vehicles heading to an intersection, a solver needs to check all possible scenarios; the solver needs to undergo a search process to find the optimum solution among all possible scenarios. The complexity of finding an optimum solution increases when multiple factors, such as vehicles at different approaches, pedestrians, and cyclists, are considered. In this case, quantum computers could play a promising role. Unlike classical computers, quantum computers work with the laws of quantum mechanics. The basic unit of information in a quantum computer is a quantum bit or qubit. A qubit can stay in a 0-state, 1-state, or a state between 0 and 1, called a superposition [5]. An n-bit classical system can store only one combination at a time out of $2^n$ combinations, whereas an n-qubit quantum system can represent a superposition of all $2^n$ combinations simultaneously. Since a single qubit can represent multiple states, a quantum computer can process multiple states in parallel compared to a classical computer. While the processing speed for a conventional computer increases linearly with the number of bits (n), the processing speed of a quantum computer increases exponentially by a factor of $2^n$. That is why quantum computing started gaining popularity in the fields of optimization [22], artificial intelligence (AI) [15], cryptography [19], communication [19] etc., to solve complex real-world problems.

For optimization problems, a quantum annealer can effectively find the best decision among all possible solution candidates, under the ideal noiseless scenario, because a quantum annealer undergoes a mechanism called quantum annealing, which is designed to determine the best solution (i.e., the global minimum) to a given optimization problem. Quantum annealing works by using quantum fluctuations to guide the search for the best solution, and it has been shown to be effective for a wide range of optimization problems [3], such as in operational research [39], finance [36], transport optimization [7], and medical science [47].

Before sending any optimization problem to a quantum annealer, like the D-Wave machine, it is necessary to formulate the problem mathematically as a Binary Quadratic Model (BQM). There are two types of BQM models to formulate any problem for a D-Wave annealer: the use of the Ising model [27] and the use of the QUBO problem [31]. No matter which BQM we use, it is converted to the Ising model by the D-Wave quantum annealer before performing quantum annealing. In our case, we use the QUBO formulation. For a D-Wave quantum annealer, the objective function, which is a mathematically formulated problem to be optimized, gives an energy value as a function of variables. In a QUBO model, the variables are mapped to qubits, aiming to get the states of the qubits for which the objective function remains at the lowest energy value. Some problems have multiple minimum points in their energy landscape, where the lowest point is known as the global minimum. The other minimum points are called local minima, as shown in Fig. 1. Fig. 1 is a simplified representation of an energy landscape that is shown in two dimensions for the ease of explanation. However, an actual energy landscape has more than two dimensions. When a classical optimizer tries to determine the global minimum, if it is stuck at one of the local minima, depending on the initial point it starts from, it could consider it as the optimal solution of an objective function. To find the global minimum, a classical algorithm needs to pass through the energy barrier, as shown in Fig. 1. The presence of quantum tunneling in a quantum annealer enables pathways among different parameters in these minima without going through the barriers between them. Such a tunneling process in a quantum annealer allows us to explore larger energy landscapes without traversing the entire parameter space [24]. Quantum annealers like D-Wave machines, though with limited tunneling channels, can sufficiently increase the chances of finding the global minimum, as shown in Fig. 1. Once the optimal energy states are found, the states of the qubits are measured and converted into the corresponding solution space with respect to the variables originally considered in the objective function. This is how quantum annealers can outperform classical computers in terms of optimization. The efficacy of a quantum annealer motivated us to utilize it in our VTL application.

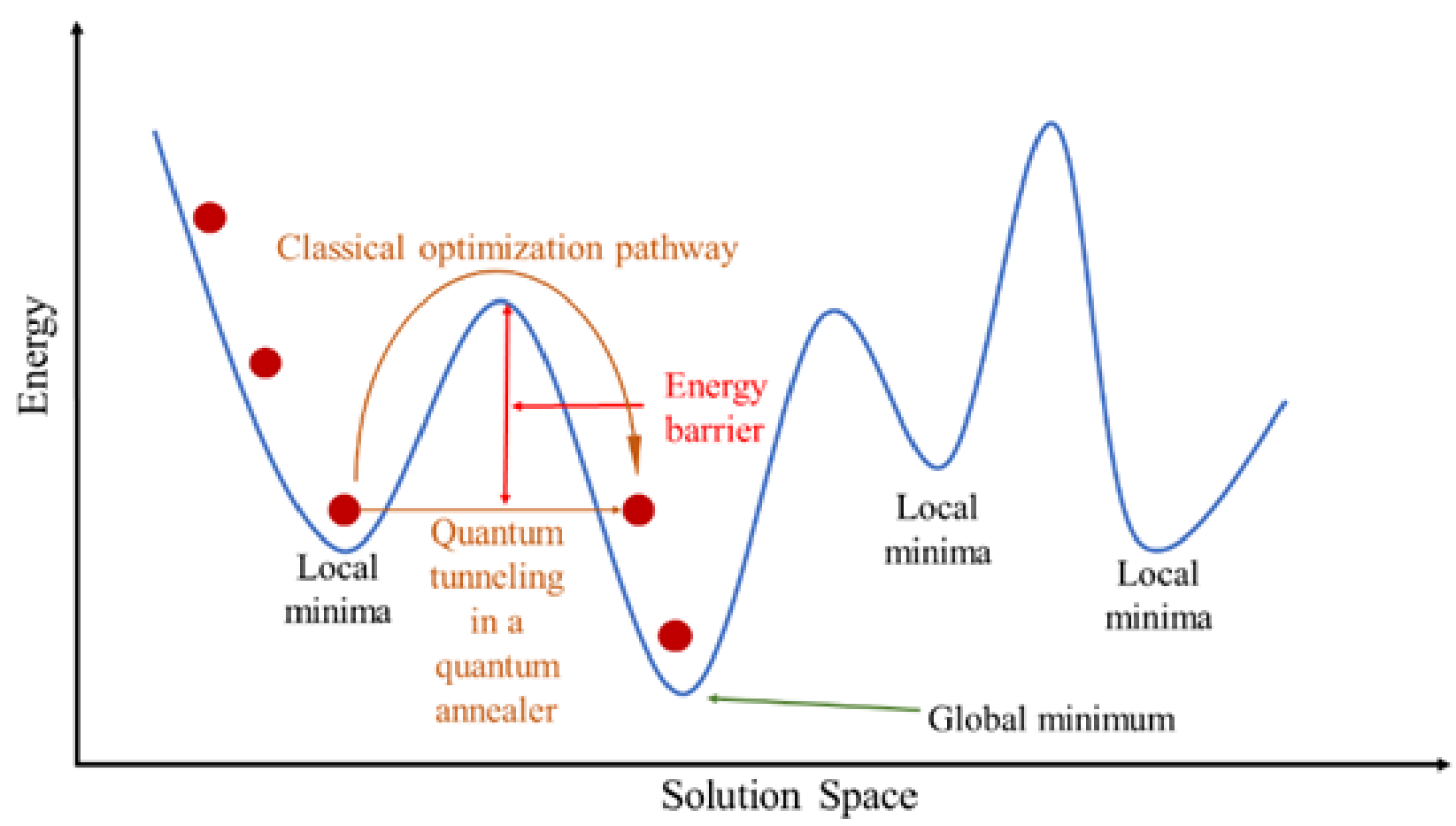

Figure 1: Quantum tunneling to explore an energy landscape to search for the global minimum (adapted from [46]).

## 4 QUANTUM ANNEALING-SUPPORTED VTL METHOD

The VTL problem in this study can be considered as an Asymmetric Open-Loop Traveling Salesman Problem (AOTSP). A Traveling Salesman Problem (TSP) is an optimization problem that involves finding the shortest route while visiting each city only once among the given cities and finally returning to the starting city [32]. In an Asymmetric Traveling Salesman Problem (ATSP), the distances between the cities are asymmetric, meaning the distances between the two cities differ depending on the direction of travel [35]. An Open-Loop Traveling Salesman Problem (OTSP) involves visiting each city only once, but not returning to the starting city [10]. Hence, AOTSP is an optimization problem that involves finding the shortest route while visiting each city only once among the given cities, where the distances between two cities are asymmetric, without returning to the starting city.

In the context of traffic lights, the National Electrical Manufacturers Association (NEMA) phase diagram provides a standardized representation of traffic signal phases at an intersection [51]. In this study, we choose the standard 8 NEMA phases while excluding exclusive right-turns, considering right-turning vehicles will move with the through traffic, as shown in Fig. 2(a). In our case study, there are 8 different movements or phases at an intersection, as shown in Fig. 2(b). These 8 phases in VTL can be considered as cities in the AOTSP, where each conflicting movement of Fig. 2(b) must be given the right-of-way one by one, just like visiting each city, and the movements that have already been given right-of-way will not be given the right-of-way again in the same signal cycle just like each city is visited only once and the starting city will not be revisited in AOTSP. The stopped delays for the CVs in one phase, caused by CVs in another phase being given the right-of-way immediately beforehand, are analogous to the distances between two cities in the AOTSP.

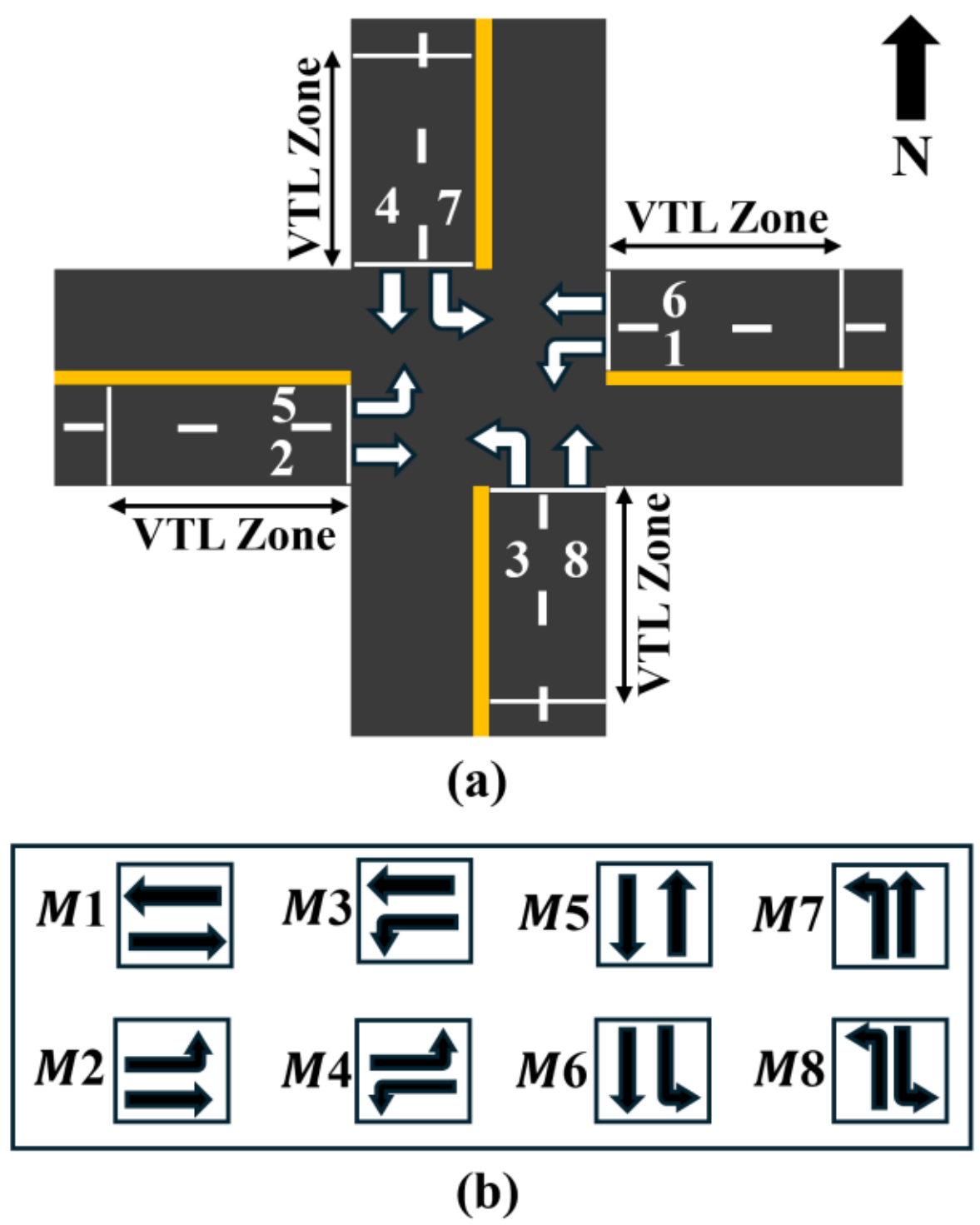

Figure 2: NEMA phase diagram showing (a) NEMA phase notation and (b) 8 unique sets of movements (M), phases, considered in this study.

Let us assume that the speeds and positions of the CVs within a VTL zone as they approach an intersection are known. The length of the VTL zone for each approach is defined as the distance to an upstream point on that approach from the stop line, within which all CVs are considered in a VTL application for optimal traffic flow. For each CV, the estimated time of arrival (ETA) can be calculated by dividing the distance to the intersection by its speed. For the CVs that have already arrived at the intersection and with a speed of zero, the ETA is considered zero. Using the ETAs, the stopped delays of CVs between two consecutive phases can be determined. For example, consider that we want to calculate the stopped delays for the CVs in phase 6 caused by the CVs in phase 5, which is given the right-of-way immediately before phase 6 in Fig. 2. Then, first, the furthest CVs within the VTL zone for northbound through (NBT) and southbound through (SBT) (phase 5 in Fig. 2) approaches are identified, and their ETAs are calculated. Among these two CVs, the one with the higher ETA is considered the last CV of phase 5. The stopped delays for all CVs in phase 6 are then measured relative to the last CV of phase 5. If the ETA of the last CV of phase 5 is $t_5$, and the ETA of the i$^{th}$ CV in phase 6 is $t_{6i}$, then the total stopped delay $d_{56}$ for the CVs in phase 6 caused by phase 5 CVs can be calculated as:

$$d_{56} = \sum_{i} max(0, t_5 - t_{6i} + Y + R) \tag{1}$$

where, $Y$ and $R$ represent the yellow and red-light intervals, respectively.

Similarly, the total stopped delays of the CVs in one phase, relative to CVs in another phase that was given the right-of-way immediately beforehand, can be calculated. These delays correspond to the distances between cities in the AOTSP. This is how the VTL can be mapped to an 8-city (equivalent to 8 NEMA phases) AOTSP problem, where the objective is to determine the optimal phase sequence for the CVs that minimizes the total stopped delay for all the approaching vehicles within the VTL zone. Finding the optimal route in the AOTSP, equivalent to identifying the optimal phase sequence for the CVs from all possible combinations, is a combinatorial optimization problem [37]. Previous studies indicate that quantum optimization delivers more promising results than classical optimization for solving combinatorial optimization problems [4, 45].

The mathematical formulation or the objective function of our problem of interest can be formulated as a QUBO problem or an Ising model. A QUBO problem can be represented by the following equation [31]:

$$f(\boldsymbol{x}) = \sum_{i} Q_{ii} x_i + \sum_{i<j} Q_{ij} x_i x_j \tag{2}$$

where, $\sum_i Q_{ii} x_i$ are the linear and $\sum_{i<j} Q_{ij} x_i x_j$ are the quadratic terms, $\boldsymbol{x} = x_1, x_2, ...$, such that $x_i \in \{0,1\}$, and $Q_{ii}$ and $Q_{ij}$ are the associated weights of the linear and the quadratic terms, respectively.

To formulate the objective function of the VTL application as an AOTSP, which is compatible with solving on a quantum annealer, we use QUBO. The VTL problem can be formulated by equations (3)-(5).

$$\min_{x_{i,k}} [\sum_{i=1}^{n} \sum_{j=1, j \neq i}^{n} d_{ij} \sum_{k=1}^{n-1} x_{i,k} x_{j,k+1}] \tag{3}$$

$$\sum_{k=1}^{n} x_{i,k} = 1 \tag{4}$$

$$\sum_{i=1}^{n} x_{i,k} = 1 \tag{5}$$

where, $x_{i,k}$ is a binary variable representing whether city $i$ is visited at position $k$ in the sequence, which is equivalent to phase $i$ being at position $k$ in the phase sequence. $d_{ij}$ represents the distance from city $i$ to city $j$, which corresponds to the total stopped delay for the CVs in phase $j$, caused by the CVs in phase $i$ being given the right-of-way immediately before phase $j$, can be measured using equation (1). The total number of possible phases is $n = 8$, as there are eight phases in total.

From equations (3)-(5), we find that there are a total of 64 binary decision variables, where each variable represents a phase and its position in the phase sequence. If the optimizer's solution indicates that a variable $x_{i,k} = 1$, it means the CVs in phase $i$ will be given the right-of-way in the $k^{th}$ position of the phase sequence to minimize the total delay. For 64 binary variables, there are a total of $2^{64}$ possible choices from which the optimizer has to determine the best option for which the total delay is the minimum. Equation (4) ensures that each city $x_i$ is visited exactly once in the AOTSP, which means each phase is given exactly one position in the phase sequence (exactly one variable from $x_{i,1}$, $x_{i,2}$, $x_{i,3}$, $x_{i,4}$, $x_{i,5}$, $x_{i,6}$, $x_{i,7}$, and $x_{i,8}$ can be 1 at a time). Equation 5 ensures that each position on the tour is occupied by exactly one city in the AOTSP, which means each position of the phase sequence is given to exactly one phase (exactly one variable from $x_{1,k}$, $x_{2,k}$, $x_{3,k}$, $x_{4,k}$, $x_{5,k}$, $x_{6,k}$, $x_{7,k}$, and $x_{8,k}$ can be 1).

The QUBO formulation of the objective function for the VTL problem, considering all the conditions in equations (3)-(5), is presented in equation (6):

$$\min_{x_{i,k}}\left[\left(\sum_{i=1}^{n}\sum_{j=1,j\neq i}^{n} d_{ij}\sum_{k=1}^{n-1} x_{i,k}x_{j,k+1}\right)+\gamma\left(\sum_{k=1}^{n} x_{i,k}-1\right)^2+\gamma\left(\sum_{i=1}^{n} x_{i,k}-1\right)^2\right] \tag{6}$$

where $\gamma$ is the Lagrange parameter, included as a multiplier with the conditions in equations (4)-(5) to ensure that these conditions are satisfied. If a condition is not met, the corresponding term adds a penalty, increasing the value of the objective function. This, in turn, reflects an increase in the waiting time or delay for vehicles. To enforce this, the value of $\gamma$ is set to 100 to impose a high penalty, although any positive number could be used. In our method, if no CVs are approaching the intersection from one or more phases, the problem is treated as an $(n-m)$-city AOTSP, where $m$ represents the number of unoccupied phases.

The quantum annealer converts the QUBO problem into an Ising model. The Ising model can be represented by a linear transformation of the binary variables $x_i = (1 + s_i)/2$. The generalized Ising model is shown in the following equation [27].

$$f(\boldsymbol{s}) = \sum_{i} h_i s_i + \sum_{i<j} J_{ij} s_i s_j \tag{7}$$

where, $\sum_i h_i s_i$ and $\sum_{i<j} J_{ij} s_i s_j$ are the linear and quadratic terms, respectively, $\boldsymbol{s} = s_1, s_2, \ldots$ represents the spins, such that $s_i \in \{-1, +1\}$ where +1 and -1 represent upward and downward spins, respectively, and can decode the two states in a qubit in general, and $h_i$ and $J_{ij}$ are the associated weights of the linear and the quadratic terms, respectively. In some quantum systems, such as in a D-Wave system, $h_i$ and $J_{ij}$ are called qubit bias and coupling strength, respectively.

The mathematical formulation of the Ising model is called the Ising Hamiltonian, which represents the energy of a spin system. The objective is to minimize the Ising Hamiltonian. The Ising model is then transformed to fit the physical chip. By this conversion to the Ising Hamiltonian, the VTL optimization problem is encoded by qubits in the quantum annealer. This conversion enables the annealer to undergo quantum annealing, aiming to find the spin alignments of the qubits for which the Hamiltonian's energy state is the lowest. At the end of the quantum annealing, the readout process collapses the state of each qubit into one of the given bases (+1 and -1). The result is then translated into 0 for -1 and 1 for +1. Based on the results of the decision variables $x_{i,j}$, the first phase ($x_{i,1}$), which represents the phase in the first position of the optimized phase sequence, is selected, and the CVs of that phase within the VTL zone are given the right-of-way. Once all the CVs of phase $x_{i,1}$, that are within the VTL zone, clear the intersection, the newly arrived CVs of that phase are given yellow and then red lights. After the CVs of the first phase have cleared, the traffic scenario changes, so the optimizer is called again to re-optimize, considering all the concurrent phases occupied with CVs. The VTL application then follows the same process.

## 5 EVALUATION STRATEGY

### 5.1 A Quantum-in-the-Loop Simulation Testbed

We develop a simulation testbed to evaluate our quantum annealing-based VTL application. The testbed connects the CVs in the SUMO (roadway traffic simulator) with a cloud-based D-wave quantum computer, as shown in Fig. 3. When the simulation starts, CVs begin moving toward the intersection. SUMO features a Python-based interactive Traffic Control Interface (TraCI) that allows CVs to be controlled in real-time. Using TraCI, BSMs, such as position and speed, are extracted. From the extracted BSMs, the CVs' current lane, distances from the intersection stop line, and speeds are used to calculate their ETAs. This data is then utilized to formulate the QUBO objective function within TraCI. The integration of SUMO with the D-Wave cloud-supported quantum annealer is achieved through TraCI. The QUBO formulation is sent to D-Wave for optimization via TraCI. Once the quantum annealing process is complete, D-Wave sends the optimized results back through TraCI. Based on these results, CVs are given the right-of-way, and they maneuver through the VTL-controlled intersection according to the instructions received from the cloud.

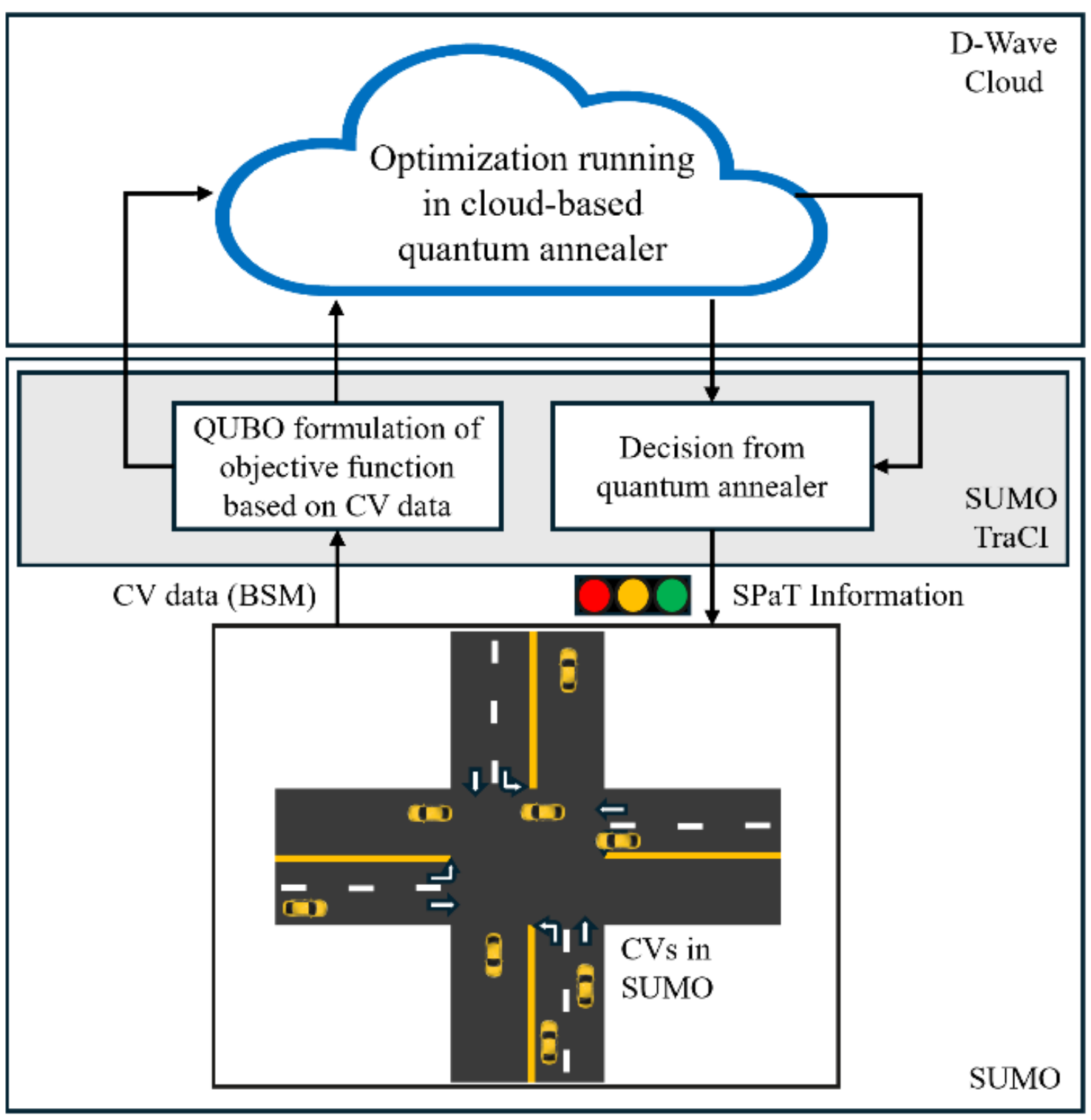

Figure 3: Simulation testbed architecture.

### 5.2 Simulation Settings

To implement and evaluate our VTL method, we created a 4-way intersection in SUMO. Each approach of the intersection has two lanes: the right lane is designated for through and right-turn movements, while the left lane is dedicated to left turns. The speed limit is set to 35 miles per hour (mph) (approximately 56 kilometers per hour). CVs are generated considering varying traffic volumes—35%, 70%, and 105% of capacity—where full capacity (100%) is defined as 1,800 passenger cars per hour per lane (pc/hr/lane) per hour of green. The base saturation flow rate is 1,900 pc/hr/ln under ideal conditions, as specified in the Highway Capacity Manual. However, the capacity of a four-way signalized intersection is typically lower than this saturation flow rate and can vary up to 1,800 pc/hr/ln due to various factors, such as the presence of heavy vehicles, roadway geometry, and the number of lanes [38]. For this case study, a capacity of 1,800 pc/hr/ln per hour of green time is assumed for each approach lane of the intersection. SUMO provides a mechanism to generate traffic over a given time interval, which can be adjusted to simulate different traffic volumes. For 35% capacity, 630 CVs are generated per lane per hour, with a time interval of 5.71 seconds/vehicle. Similarly, for 70% and 105% capacities, the time intervals are 2.86 seconds and 1.90 seconds per lane, respectively. Passenger cars are considered for all CVs. Under each traffic condition, simulations are run for assumed VTL zones of lengths 50, 75, and 100 meters to assess the consistency of quantum annealing across different lengths of VTL zones.

### 5.3 Baseline Methods

In our VTL application, optimization is performed using the D-Wave quantum annealer. To evaluate the performance of this quantum optimization, we also optimize the VTL application using two classical optimization methods: Binary Hill Climbing, and Simulated Annealing. The results are then compared to assess the effectiveness of the quantum approach.

Optimization strategies considered in this study, Binary Hill Climbing and Simulated Annealing, are appropriate for various facets of complex solution spaces having several local minima and a global minimum. While it can occasionally be stuck in deep local minima, Binary Hill Climbing, a variety of Hill Climbing where the variables are binary, is an effective method of exploring discrete, combinatorial spaces by iteratively improving solutions through neighborhood searches, where neighboring solutions are generated by flipping one or more bits in the current binary solution [53]. This allows it to avoid shallow local minima and incrementally move toward better solutions. The Hill Climbing method is widely used to search for optimum solutions, such as in the Bayesian network structure learning algorithm [37, 50] etc., due to its efficacy in solving complex optimization problems. Comparably, Simulated Annealing is a probabilistic technique inspired by the annealing process in metallurgy. It is effective in both discrete and continuous solution spaces and is particularly useful for escaping local minima by allowing occasional uphill moves based on a temperature parameter that gradually decreases over time [43]. This temperature schedule helps balance exploration and exploitation, improving the chances of locating the global minimum. Simulated Annealing has been widely used in solving complex optimization problems, especially where the solution space is rugged and non-convex. We use the Simulated Annealing algorithm provided by D-Wave, which serves as a classical benchmark for comparison against Quantum Annealing. This comparison, between Quantum Annealing and Simulated Annealing, is fair because both solve the same class of optimization problems, such as QUBO or Ising models. Both use analogous stochastic strategies (Simulated Annealing via thermal fluctuations and Quantum Annealing via quantum tunneling), making Simulated Annealing a fair classical baseline [8, 33, 42].

In addition, we compare our method with traditional roadway infrastructure-based traffic control systems. We consider an actuated traffic signal as a baseline traditional traffic control system to compare performance with our VTL method. An actuated traffic signal assigns right-of-way depending on the presence of vehicles or pedestrians, adjusting its operation based on real-time demand instead of adhering to a fixed time cycle.

## 6 EVALUATION OUTCOMES

For the performance evaluation, we consider vehicle waiting times or delays at the intersection and travel times through the simulated network. Delay, in this study, is defined as the total duration during which a vehicle's speed is below 0.1 meters per second, while travel time is the time taken by a CV to travel through the roadway section we simulated [48]. To highlight the performance of the optimization methods, we also evaluate the cost function values achieved by the optimizers during the optimization process.

### 6.1 Performance Comparison with Classical Optimizers

*6.1.1CV Delay.*

We evaluate the VTL application under varying traffic volumes and different lengths of the VTL zone. The delay results, presented in Fig. 4, demonstrate that quantum annealing consistently reduces delays for CVs compared to classical optimization methods across various traffic volumes and the length of VTL zones. For a 95% confidence interval, the overall average delay per vehicle, considering all traffic volumes and the length of VTL zones, is 111.3 seconds for Hill Climbing, 91.1 seconds for Simulated Annealing, and 46 seconds for Quantum Annealing. The average and maximum vehicle delays are lower with quantum annealing, as shown in Fig. 4.

To statistically evaluate these findings, we conducted statistical two-sample t-tests at a significance level of 0.05, comparing the average delays for quantum annealing-based VTL with those of classical methods across different traffic volumes and VTL zones of different lengths. The hypotheses for these tests are as follows:

- Null Hypothesis ($H_0$): The mean delay of CVs using quantum annealing-based VTL is not significantly lower than (i.e., is equal to or greater than) the mean delay using classical methods.
- Alternative Hypothesis ($H_a$): The mean delay of CVs using quantum annealing-based VTL is significantly lower than that using classical methods.

We performed two-sample, one-tailed t-tests for unequal variances (as variances are found unequal in our results) to compare the delays for Quantum Annealing against each classical method across various traffic volumes and VTL zones of different lengths. The results, summarized in Table 1, show $p < 0.05$ in all cases, except for Simulated Annealing under a traffic volume of 35% of capacity, where QA, SA, and HC are used as the abbreviations of Quantum Annealing, Simulated Annealing, and Hill Climbing, respectively. These results indicate that the null hypothesis can be rejected under all conditions except one. In the case of 35% traffic volume, although the t-test does not confirm that the stopped delay of CVs for Quantum Annealing-supported VTL is significantly lower than Simulated Annealing, the average stopped delay is still lower for Quantum Annealing. In all other scenarios, the stopped delays of CVs are significantly lower for the quantum annealing-based VTL compared to the classical methods at a 0.05 significance level. Therefore, we conclude that Quantum Annealing significantly reduces CV delays in comparison to classical optimization methods across most tested conditions.

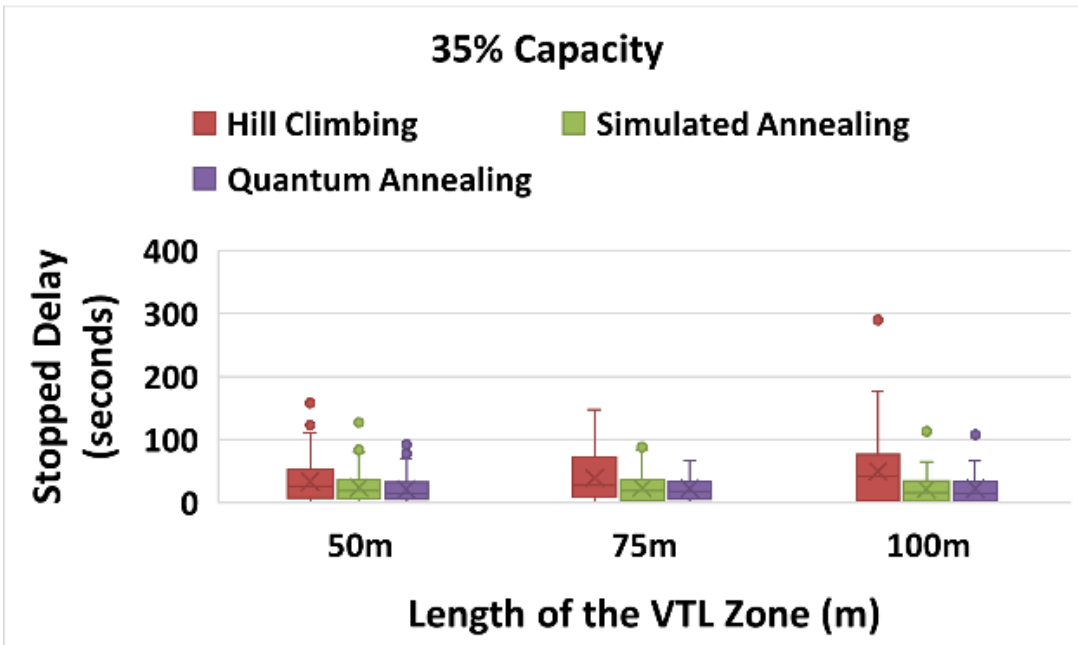

**(a)**

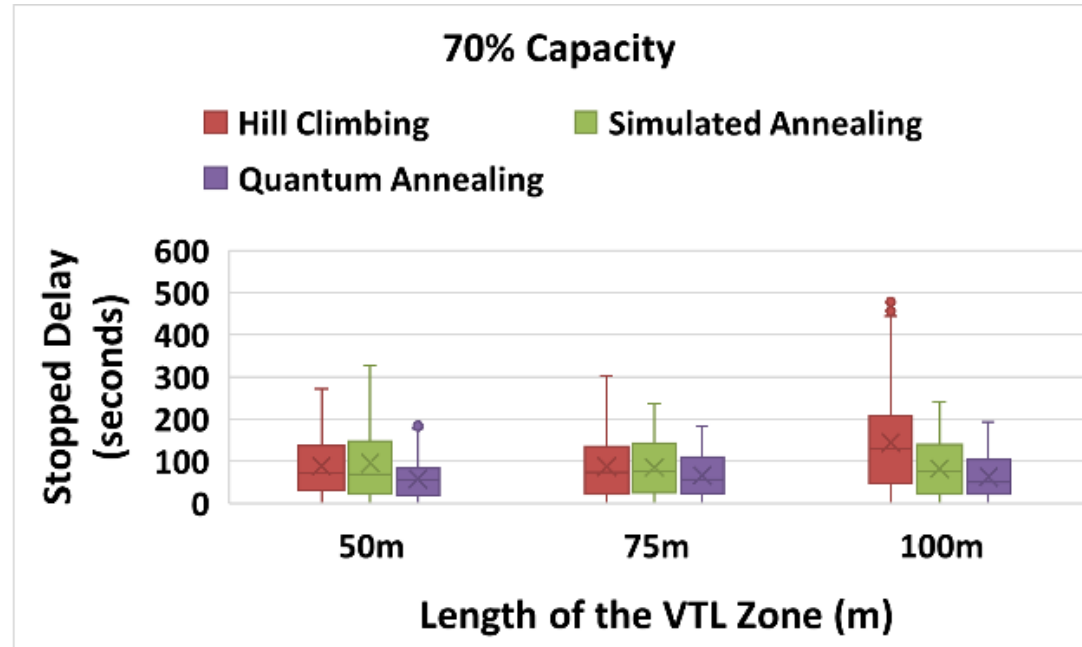

**(b)**

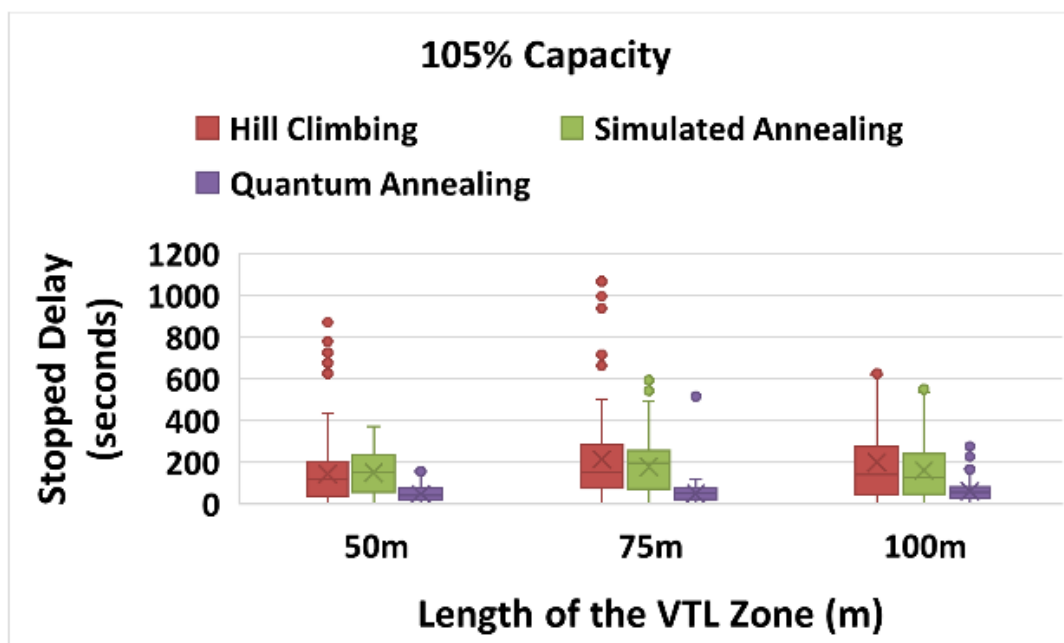

**(c)**

Figure 4: Delay of CVs for different optimization methods under (a) 35%, (b) 70%, and (c) 105% traffic capacity.

Table 1: p-values of two-sample t-tests for delays of various lengths of the VTL zone and traffic volumes.

| Two-sample t-test | VTL Zone (m) | 35% Capacity | 70% Capacity | 105% Capacity |
|---|---|---|---|---|
| $H_o$: $\mu_{QA} \geq \mu_{HC}$ | 50 | $2.3\times10^{-5}$ | $4.0\times10^{-11}$ | $1.3\times10^{-39}$ |
| $H_a$: $\mu_{QA} < \mu_{HC}$ | 75 | $1.9\times10^{-6}$ | $1.5\times10^{-5}$ | $1.8\times10^{-46}$ |
| | 100 | $6.2\times10^{-9}$ | $5.0\times10^{-25}$ | $1.7\times10^{-43}$ |
| $H_o$: $\mu_{QA} \geq \mu_{SA}$ | 50 | 0.1 | $5.9\times10^{-12}$ | $3.7\times10^{-79}$ |
| $H_a$: $\mu_{QA} < \mu_{SA}$ | 75 | 0.5 | $4.5\times10^{-5}$ | $9.1\times10^{-76}$ |
| | 100 | 0.42 | $9.7\times10^{-6}$ | $1.6\times10^{-45}$ |

*6.1.2CV Travel Time.*

Fig. 5 demonstrates that quantum annealing consistently reduces travel times for CVs compared to classical optimization methods across various traffic volumes and VTL zones of different lengths. For a 95% confidence interval, the overall travel time per vehicle, considering all traffic volumes and lengths of the VTL zone, is 181.3 seconds for Hill Climbing, 161.7 seconds for simulated annealing, and 115.3 seconds for Quantum Annealing. The average and maximum travel times are lower with quantum annealing, as shown in Fig. 5.

To statistically evaluate these findings, we again conducted statistical two-sample t-tests at a significance level of 0.05, comparing the average delays for quantum annealing-based VTL with those of classical methods across different traffic volumes and VTL zones of different lengths. The hypotheses for these tests are as follows:

- Null Hypothesis ($H_0$): The mean travel time of CVs using quantum annealing-based VTL is not significantly lower than (i.e., is equal to or greater than) the mean travel time using classical methods.
- Alternative Hypothesis ($H_a$): The mean travel time of CVs using quantum annealing-based VTL is significantly lower than that using classical methods.

We performed two-sample, one-tailed t-tests for unequal variances (as variances are found unequal in our results) to compare the travel times for quantum annealing against each classical method across various traffic volumes and VTL zones of different lengths, except for Simulated Annealing under a traffic volume of 35% of capacity. These results indicate that the null hypothesis can be rejected under all conditions except one. In the case of 35% traffic volume, although the t-test does not confirm that the travel time of CVs for Quantum Annealing-based VTL is significantly lower than Simulated Annealing, the average travel time is still lower for Quantum Annealing. In all other scenarios, the travel times of CVs are significantly lower for the quantum annealing-based VTL compared to the classical methods at a 0.05 significance level. Therefore, we conclude that Quantum Annealing significantly reduces CV travel times in comparison to classical optimization methods across most tested conditions.

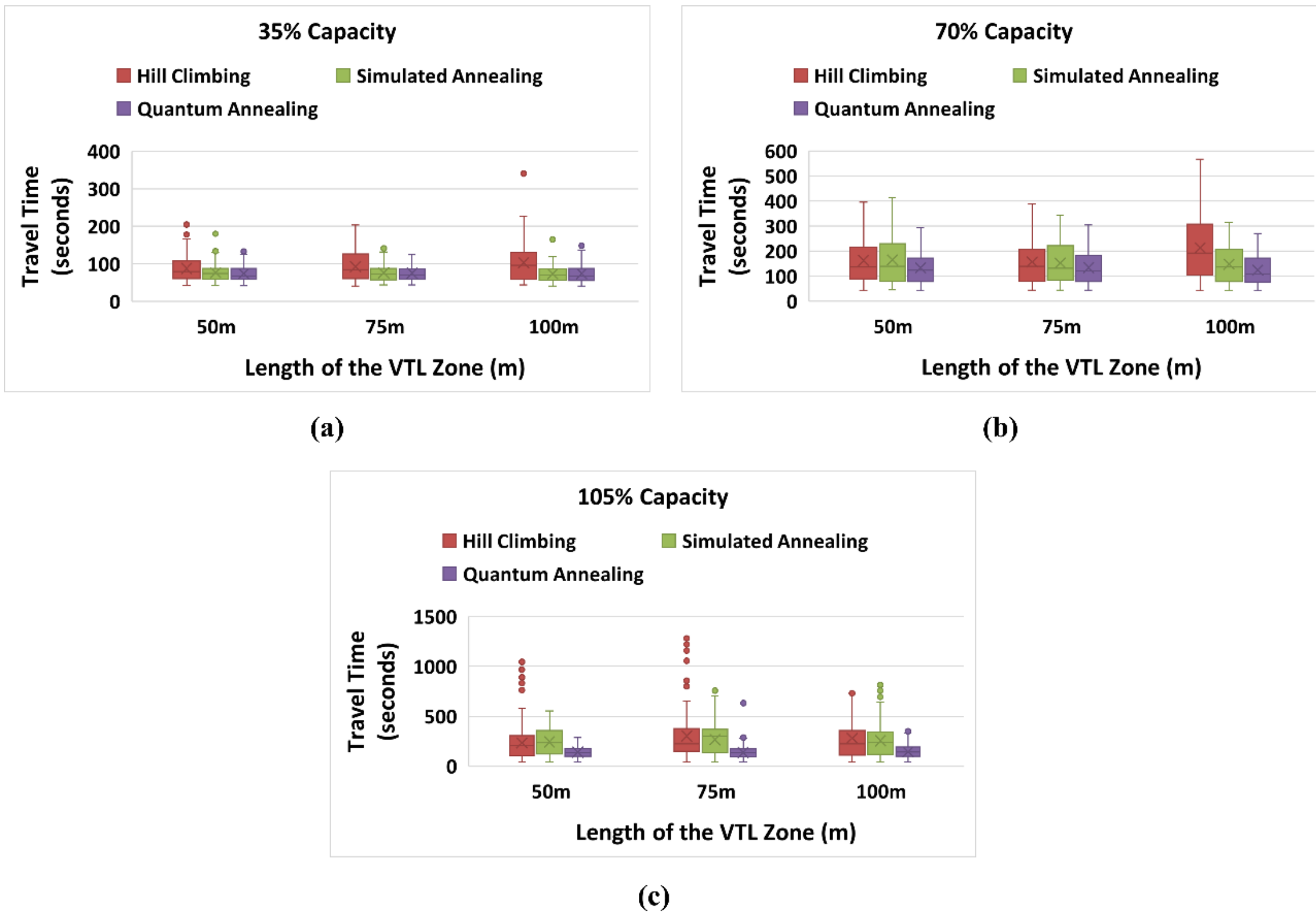

Figure 5: Travel time of CVs for different optimization methods under (a) 35%, (b) 70%, and (c) 105% traffic capacity.

Table 2: p-values of two-sample t-tests for travel time across various lengths of the VTL zone and traffic volumes.

| Two-sample t-test | VTL Zone (m) | 35% Capacity | 70% Capacity | 105% Capacity |
|---|---|---|---|---|
| $H_o$: $\mu_{QA} \geq \mu_{HC}$ | 50 | $1.3\times10^{-5}$ | $9.0\times10^{-7}$ | $1.4\times10^{-28}$ |
| $H_a$: $\mu_{QA} < \mu_{HC}$ | 75 | $6.5\times10^{-7}$ | $9.8\times10^{-5}$ | $1.1\times10^{-39}$ |
| | 100 | $2.8\times10^{-9}$ | $1.3\times10^{-24}$ | $4.5\times10^{-37}$ |
| $H_o$: $\mu_{QA} \geq \mu_{SA}$ | 50 | 0.21 | $5.1\times10^{-7}$ | $2.4\times10^{-47}$ |
| $H_a$: $\mu_{QA} < \mu_{SA}$ | 75 | 0.38 | $6.4\times10^{-4}$ | $2.4\times10^{-59}$ |
| | 100 | 0.42 | $7.6\times10^{-6}$ | $2.9\times10^{-37}$ |

*6.1.3Cost Function Evaluation.*

The values of the cost function after each optimization run are measured for all optimization methods mentioned under all conditions. Fig. 6 presents the cost function values after optimization for all the mentioned methods. A lower cost function value indicates a more optimal outcome, as our goal is to minimize vehicle delays. From Fig. 6, it is evident that Quantum Annealing outperforms all the classical optimization methods considered in this study, with the lowest average cost function value of 0.22.

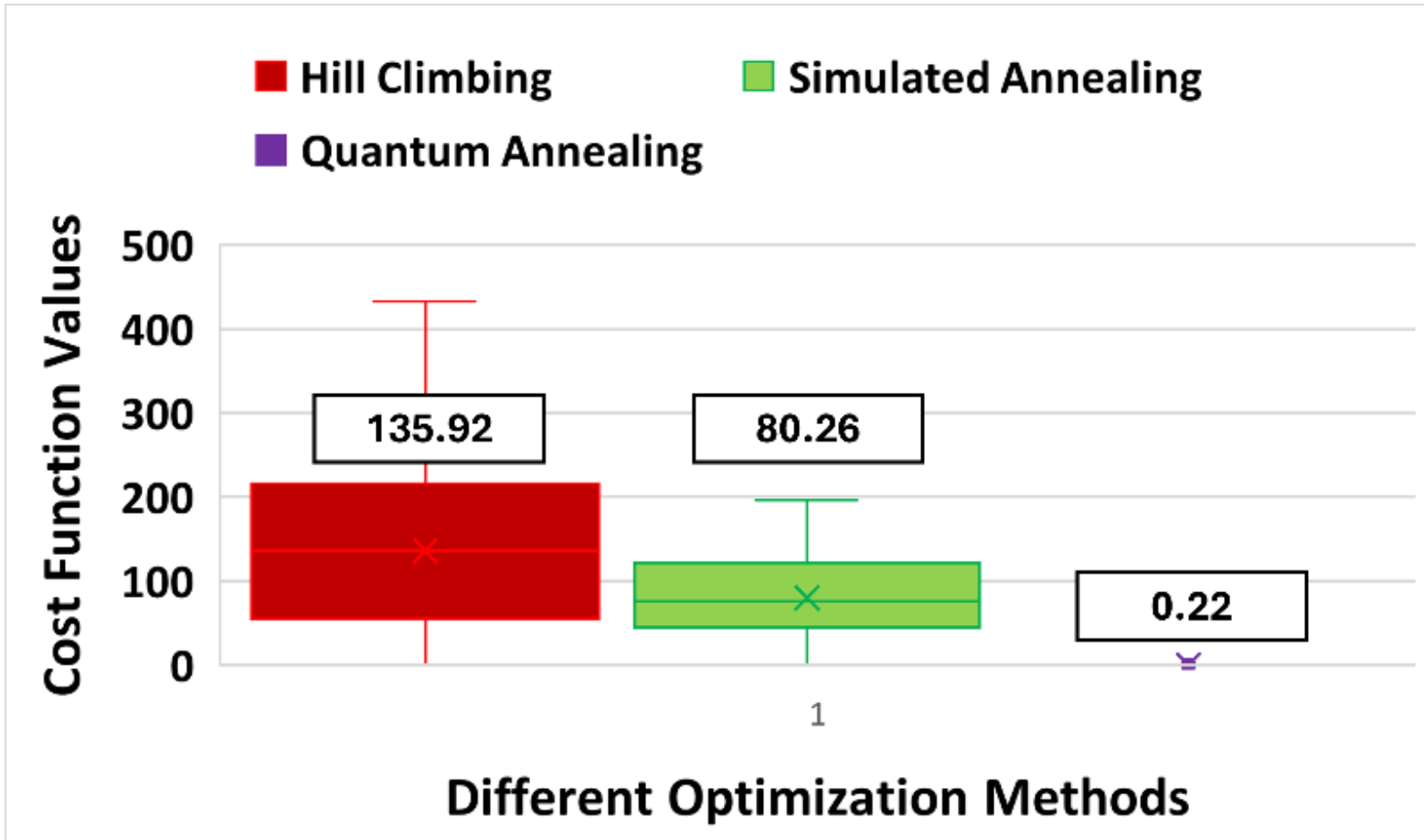

Figure 6: Cost function values for different optimization methods.

### 6.2 Performance Comparison with Traditional Traffic Signal

We also compare the performance of the Quantum Annealing-based VTL with the traditional actuated traffic control system. As a traditional traffic control system for our simulated intersection in SUMO, we choose an actuated traffic signal control system. The actuated traffic signal timing is optimized using Synchro. We obtain the optimized traffic signal timing from Synchro for different traffic volumes that we integrate in SUMO to compare vehicle stopped delays and travel times with Quantum Annealing-based VTL. The results are summarized in Table 3, which reveals that Quantum Annealing-based VTL also outperforms traditional traffic control systems in all traffic volumes considered in this study.

Table 3: Comparison of stopped delay and travel time between Quantum Annealing-based VTL and traditional actuated traffic control method.

| Traffic Volume (% of Capacity) | Stopped Delays (s) | | Travel Time (s) | |
|---|---|---|---|---|
| | Actuated Traffic Signal | QA-based VTL | Actuated Traffic Signal | QA-based VTL |
| 35% | 24.3 | 21.7 | 77.3 | 74.4 |
| 70% | 66.1 | 62.5 | 131.4 | 130.7 |
| 105% | 123.4 | 53.9 | 214.4 | 140.7 |
| **Average** | **71.3** | **46.0** | **141** | **115.3** |

### 6.3 Feasibility of Implementation

In this study, we also measured the processing time of the quantum annealer and the end-to-end latency. Fig. 7 shows that the average end-to-end latency we measured during our experiment is 2.99 seconds. This latency includes (i) data upload delays from CVs to the D-Wave cloud, (ii) the queue delays in the D-Wave cloud (i.e., the time between the data upload to the D-Wave cloud and the time the Quantum Processing Unit (QPU) begins processing), (iii) the processing time for optimizing the VTL application, referred to as QPU access time in D-Wave, which includes mapping the objective function to qubits for quantum annealing, performing the annealing, and translating the optimized results from the QPU for download to CVs, and (iv) the data download delays from the cloud back to the CVs.

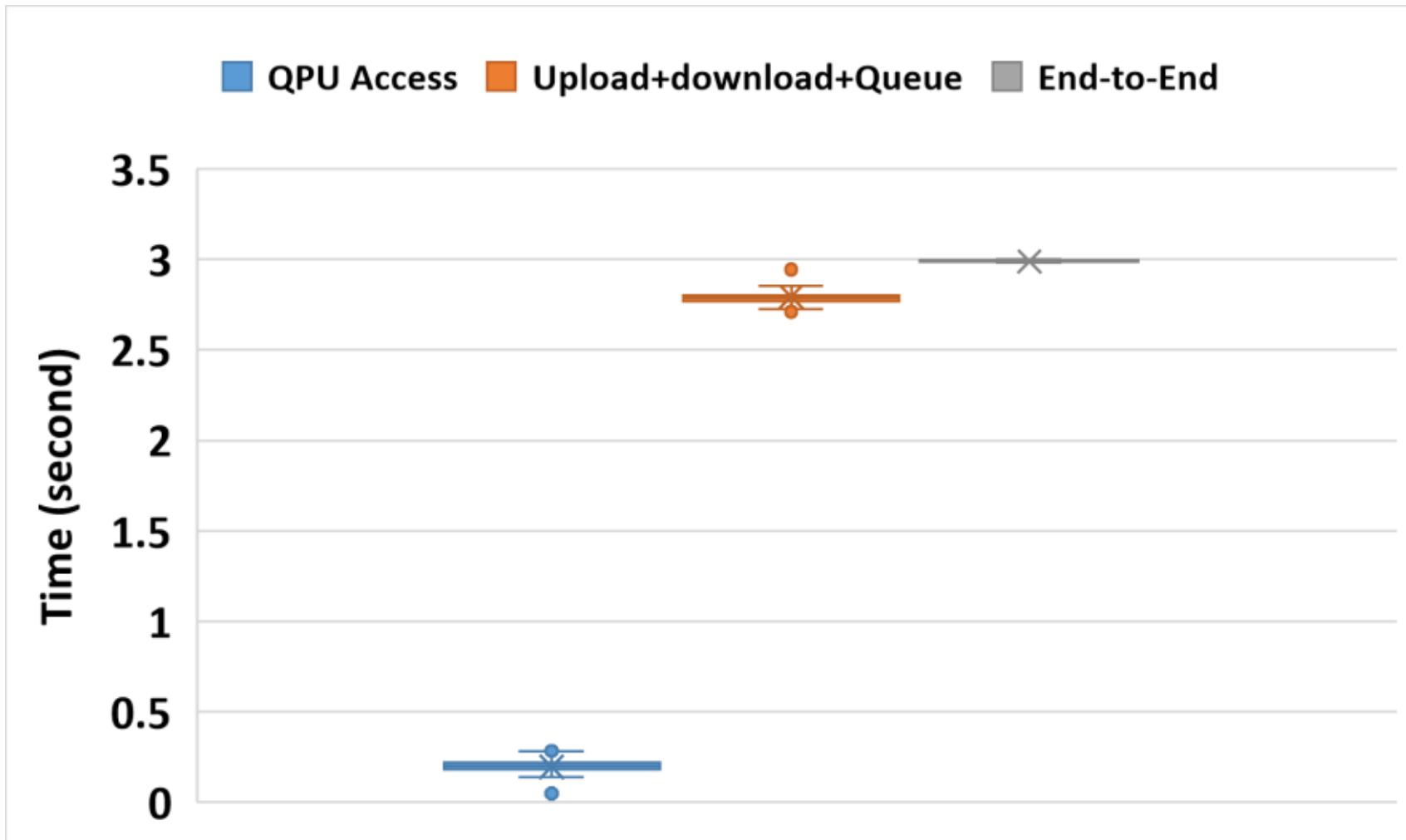

Figure 7: Latencies of VTL application in D-Wave's cloud-supported quantum infrastructure.

The threshold for end-to-end latency requirements in real-time mobility applications is 1,000 milliseconds (i.e., 1 second) [20]. However, the average end-to-end latency observed in this study is 2.99 seconds. This elevated latency is attributed to the developmental stage of quantum computers and their associated cloud infrastructures. Unlike the mature and widely accessible cloud services powered by classical computing, quantum computing infrastructures are still in the process of development and are not yet broadly available.

In this study, we utilize the D-Wave cloud-based quantum annealer to solve the optimization problem for VTL. As shown in Fig. 7, the average QPU access time (processing time) is 0.197 seconds (197 ms), which is fast enough for real-time mobility applications like VTL. However, the primary contributor to the increased end-to-end latency is not the QPU access time. Instead, the delay is mainly due to the average upload, queue, and download time, which is measured at 2.793 seconds, as shown in Fig. 7. This is because the QPU processes tasks based on its availability, which is currently limited by the restricted commercial and broad accessibility of quantum annealers.

However, based on a previous study [12] related to cloud-supported real-time CV applications at roadway intersections, we observed that the average latency for data upload, queue processing, and download in a commercially available cloud with classical computing infrastructure is approximately 169 ms. This suggests that once cloud-supported quantum infrastructures become more mature and widely accessible through different providers, similar to current cloud-based classical computing systems, our method will become feasible and effective for real-time VTL implementations.

## 7 CONCLUSIONS

In this study, we develop a quantum computing-powered VTL method for CVs at intersections. Our VTL method utilizes a cloud-supported quantum computing environment, removing the need for costly public roadway infrastructure, such as that associated with traditional traffic control systems, which require substantial investment in deployment and maintenance. VTL operates without physical traffic lights, signal controllers, or associated wired communication infrastructure that are usually deployed and maintained by public agencies. Through our cloud-based quantum infrastructure, we have observed superior outcomes compared to classical optimization-based VTL methods in terms of reducing intersection stopped delay and travel times. Our extensive analysis reveals that our quantum annealing-based VTL significantly reduces the stopped delays and travel times of CVs compared to the classical methods with a 95% confidence level across different traffic volumes and different lengths of the VTL zone. Additionally, we compare the performance of our VTL method with traditional traffic signals, and our analysis reveals that our method reduces the CV stopped delay and travel times across different traffic volumes, compared to traditional fully actuated traffic signals. Additionally, traditional traffic control systems are highly vulnerable to cyberattacks due to outdated infrastructure, lack of encryption, and weak authentication in communications with central servers, whereas quantum cloud-based platforms enhance cybersecurity through end-to-end encryption, strict access control, and secure optimization environments.

Quantum computers that exist today are still in their early stages and are limited by their capabilities. However, extensive research is ongoing to advance their computational functionalities. Recent advancements have shown significant progress, such as the development of more error-resilient quantum computers and the scaling up of qubits to boost their computational power. These developments could be employed to address more complex intersections with dedicated pedestrian movements and signals, while also considering multiple intersections to reduce stopped delays in signalized corridors

The work presented in this paper introduces a novel approach to quantum-optimized VTL systems, specifically leveraging quantum annealing. Additionally, we develop a simulation testbed integrating SUMO with a cloud-supported quantum computing platform to evaluate the theoretical VTL method we developed in this study. Our study focused on a 4-way intersection, with each approach featuring two lanes. Future endeavors will extend this study to more complex intersections, incorporating pedestrian signals, following the advancement of quantum computers. Once quantum platforms become commercially viable for real-time use with reduced cloud queue times, future work will also extend this study by implementing our method in the real world.

For real-world implementations, a roadside unit connected with the cloud-supported quantum computer and equipped with a low-latency communication radio to communicate with the CVs, such as cellular-vehicle-to-everything communication technology, will collect the vehicular data or BSMs of the CVs. Based on the CV data, the roadside unit will formulate the stopped delay minimization QUBO problem and send it to the cloud-supported quantum annealer for optimization. After the optimization problems are solved, the roadside unit will receive the optimized solution from the cloud via a secured and encrypted communication channel and then will send the SPaT info to the CVs. Additionally, considering multiple intersections and pedestrian traffic will further increase computational complexity, which will be most suitable to be solved in future quantum computers. Classical computers will struggle to provide optimum solutions in real-time in such scenarios, highlighting the promising future of quantum computing-based VTL systems.

**ACKNOWLEDGMENTS**

This work is based upon the work supported by the Center for Connected Multimodal Mobility ($C^2M^2$) (a U.S. Department of Transportation Tier 1 University Transportation Center), and the National Center for Transportation Cybersecurity and Resiliency (TraCR) (a US Department of Transportation National University Transportation Center), both headquartered at Clemson University, Clemson, South Carolina, USA. Any opinions, findings, conclusions, and recommendations expressed in this material are those of the author(s) and do not necessarily reflect the views of $C^2M^2$, TraCR, and the U.S. Government assumes no liability for the contents or use thereof.